IJCSI International Journal of Computer Science Issues, Vol. 7, Issue 1, No. 3, January 2010
ISSN (Online): 1694-0784
ISSN (Print): 1694-08141# RFID Applications: An Introductory and Exploratory Study

Kamran AHSAN[1], Hanifa SHAH[2] and Paul KINGSTON[3]

[1,2] Faculty of Computing, Engineering & Technology
Staffordshire University
Stafford, ST18 0AD, UK

[1, 3] Centre for Ageing & Mental Health**Abstract**
RFID is not a new technology and has passed through many decades of use in military, airline, library, security, healthcare, sports, animal farms and other areas. Industries use RFID for various applications such as personal/vehicle access control, departmental store security, equipment tracking, baggage, fast food establishments, logistics, etc. The enhancement in RFID technology has brought advantages that are related to resource optimization, increased efficiency within business processes, and enhanced customer care, overall improvements in business operations and healthcare. Our research is part of a big project; its aim is to produce a model for mobile technology implementation of hospital patients' movement process. However, the focus of this paper is to explore the main RFID components, i.e. the tag, antenna and reader. The results of the investigations conducted on the three RFID components will be used to develop our research model.

*Keywords:* RFID technology, RFID discovery, RFID components, RFID applications, RFID in healthcare.## 1. Introduction

RFID stands for Radio Frequency Identification and is a term that describes a system of identification [1]. RFID is based on storing and remotely retrieving information or data as it consists of RFID tag, RFID reader and back-end Database [2]. RFID tags store unique identification information of objects and communicate the tags so as to allow remote retrieval of their ID. RFID technology depends on the communication between the RFID tags and RFID readers. The range of the reader is dependent upon its operational frequency. Usually the readers have their own software running on their ROM and also, communicate with other software to manipulate these unique identified tags [3]. Basically, the application which manipulates tag deduction information for the end user, communicates with the RFID reader to get the tag information through antennas. Many researchers have addressed issues that are related to RFID reliability and capability [2]. RFID is continuing to become popular because it increases efficiency and provides better service to stakeholders [1]. RFID technology has been realized as a performance differentiator for a variety of commercial applications, but its capability is yet to be fully utilised.

## 2. RFID Evolution

RFID technology has passed through many phases over the last few decades (see figure 1). The technology has been used in tracking delivery of goods, in courier services and in baggage handling. Other applications includes automatic toll payments, departmental access control in large buildings, personal and vehicle control in a particular area, security of items which shouldn't leave the area, equipment tracking in engineering firms, hospital filing systems, etc.[4, 5]. Figure 1 shows RFID evolution over the past few decades.

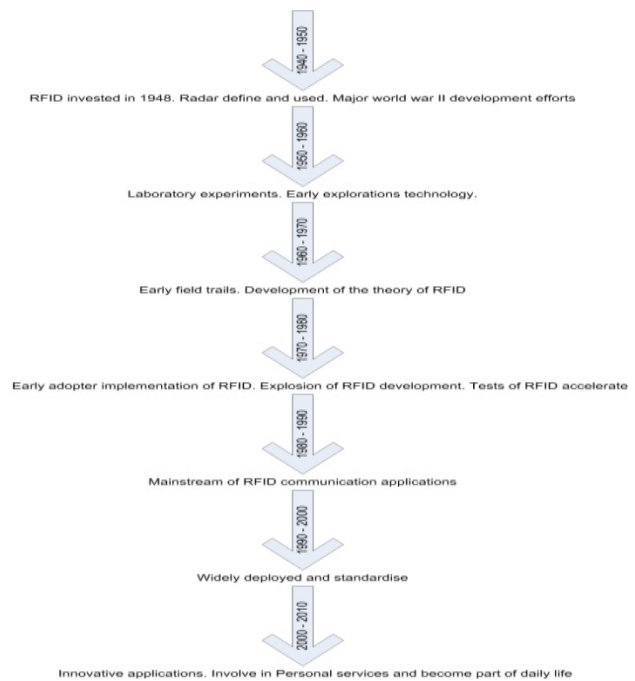

Fig. 1 RFID evolution: Over past the few decades adapted from [6]

IJCSI
www.IJCSI.org

## 3. How RFID System Works

Most RFID systems consist of tags that are attached to the objects to be identified. Each tag has its own "read-only" or "rewrite" internal memory depending on the type and application [7]. Typical configuration of this memory is to store product information, such as an object's unique ID manufactured date, etc. The RFID reader generates magnetic fields that enable the RFID system to locate objects (via the tags) that are within its range [5]. The high-frequency electromagnetic energy and query signal generated by the reader triggers the tags to reply to the query; the query frequency could be up to 50 times per second. [4]. As a result communication between the main components of the system i.e. tags and reader is established [6]. As a result large quantities of data are generated. Supply chain industries control this problem by using filters that are routed to the backend information systems. In other words, in order to control this problem, software such as Savant is used. This software acts as a buffer between the Information Technology and RFID reader [6, 7].

Several protocols manage the communication process between the reader and tag. These protocols (ISO 15693 and ISO 18000-3 for HF or the ISO 18000-6, and EPC for UHF) begin the identification process when the reader is switched on. These protocol works on selected frequency bands (e.g. 860 – 915 MHz for UHF or 13.56 MHz for HF). If the reader is on and the tag arrives in the reader fields, then it automatically wakes-up and decodes the signal and replies to the reader by modulating the reader's field [7]. All the tags in the reader range may reply at the same time, in this case the reader must detect signal collision (indication of multiple tags) [6]. Signal collision is resolved by applying anti-collision algorithm which enables the reader to sort tags and select/handle each tag based on the frequency range (between 50 tags to 200 tags) and the protocol used. In this connection the reader can perform certain operations on the tags such as reading the tag's identifier number and writing data into a tag [7].

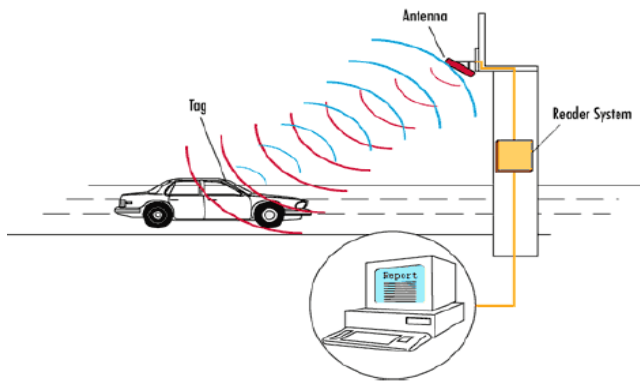

Fig. 2 A typical RFID System [7]

The reader performs these operations one by one on each tag. A typical RFID system work cycle can be seen in figure 2.

## 4. Components of an RFID System

The RFID system consists of various components which are intergrated in a manner defined in the above section. This allows the RFID system to deduct the objects (tag) and perform vaious operations on it. The intergration of RFID components enables the implementation of an RFID solution [8]. The RFID system consists of following five components (as shown in Figure 3):

- Tag (attached with an object, unique identification).
- Antenna (tag detector, creates magnetic field).
- Reader (receiver of tag information, manipulator).
- Communication infrastructure (enable reader/RFID to work through IT infrastructure).
- Application software (user database/application/ interface).

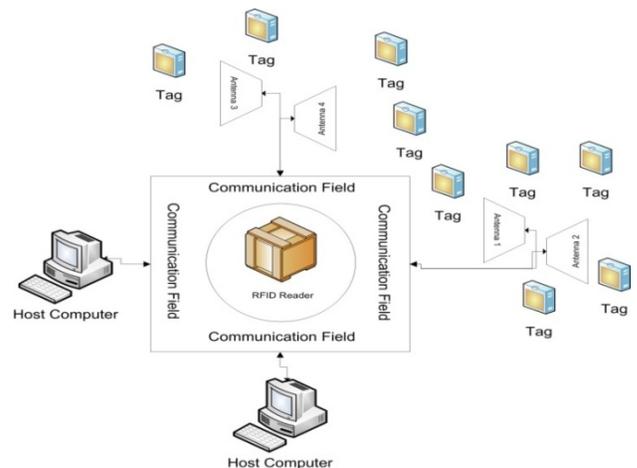

Fig. 3 Components of an RFID System

## 5. Tags

Tags contain microchips that store the unique identification (ID) of each object. The ID is a serial number stored in the RFID memory. The chip is made up of intergrated circuit and embedded in a silicon chip [7]. RFID memory chip can be permanent or changeable depending on the read/write characteristics. Read-only and rewrite circuits are different as read-only tag contain fixed data and can not be changed without re-program electonically [5]. On the other hand, re-write tags can be programmed through the reader at any time without any limit. RFID tags can be different sizes and shapes depending on the application and the environment at which it will be used. A variety of materials are intergrated on these tags. For example, in the case of the credit cards, small plastic peaces are stuck on various objects, and the labels. Labels are also emmbeded in a variety of objects such as documents, cloths, maufacturing mateirals etc [9]. Figure 4 demonstrates the different sizes and shapes of the RFID tags.

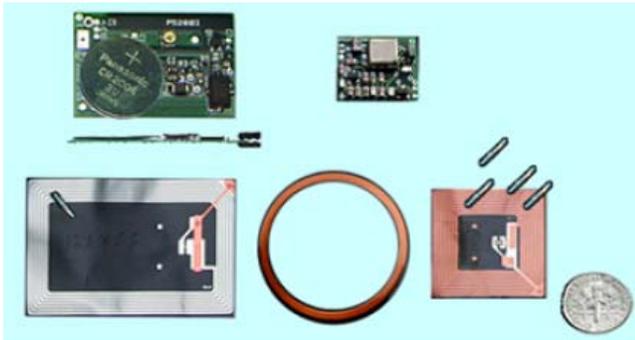

Fig. 4 Varity of RFID tags (various shape & sizes) [9]

RFID tags can also be classified by their capabilities such as read and write data [10]. Figure 5 shows the five classifications of the RFID tags [10].

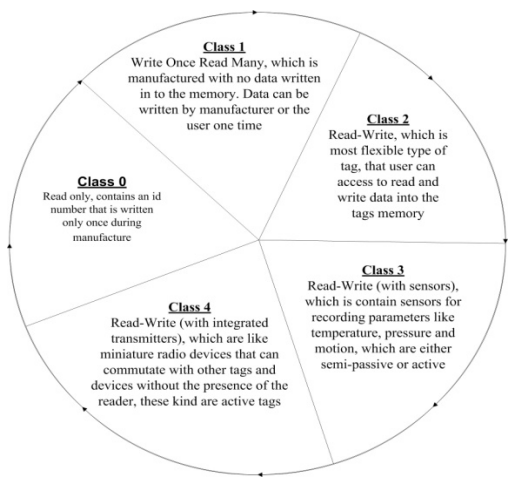

Fig. 5 RFID tags classifications

There are three types of tags: the passive, semi-active and active. Semi-active tags have a combination of active and passive tags characteristics. So, mainly two types of tags (active and passive) are being used by industry and most of the RFID system [7]. The essential characteristics of RFID tags are their function to the RFID system. This is based on their range, frequency, memory, security, type of data and other characteristics. These characteristics are core for RFID performance and differ in usefulness/support to the RFID system operations [4, 11]. While considering these characteristics, figure 6 compares the active and passive tags.

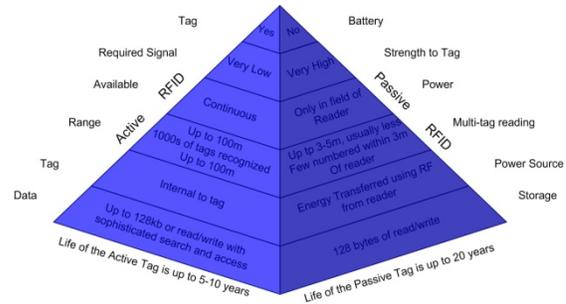

Fig. 6 RFID active and passive tags comparison

5.1 Tag Frequencies

The range of the RFID tags depends on their frequency. This frequency determines the resistance to interference and other performance attributes [12]. The use/selection of RFID tag depends on the application; different frequencies are used on different RFID tags [10]. EPCglobal and International Standards Organization (ISO) are the major organizations working to develop international standards for RFID technologies in the UHF band. These two organizations are still evolving and are not fully compatible with each other [14]. In order to avoid the use of different radio frequencies standards, most of the international communities are obligated to comply with the International Telecommunication Union (ITU) standards. The following are the commonly used frequencies:

- **Microwave** works on 2.45 GHz, it has good reader rate even faster than UHF tags. Although at this frequency the reading rate results are not the same on wet surfaces and near metals, the frequency produce better results in applications such as vehicle tracking (in and out with barriers), with approximately 1 meter of tags read range [7].

- **Ultra High Frequency** works within a range of 860-930 MHz, it can identify large numbers of tags at one time with quick multiple read rate at a given time. So, it has a considerable good reading speed. It has the same limitation as Microwave when is applied on wet surface and near metal. However, it is faster than high frequency data transfer with a reading range of 3 meters [7].

- **High Frequency** works on 13.56MHz and has less than one meter reading range but is inexpensive and useful for access control, items identifications on sales points etc as it can implanted inside thin things such as paper [6, 7].

- **Low Frequency** works on 125 kHz, it has approximately half a meter reading range and mostly used for short reading range applications

such as shops, manufacturing factories, inventory control through in and out counts, access control through showing a card to the reader. These low frequency tags are mostly not affected when applied on wet and near metal surfaces [7, 9].

## 6. Antennas

RFID antennas collect data and are used as a medium for tag reading [7]. It consists of the following:

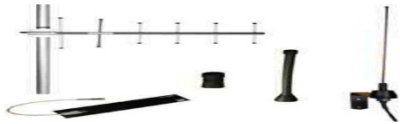

Fig. 7 RFID antennas types [11]

(1) Patch antennas, (2) Gate antennas, (3) Linear polarized, (4) Circular polarized, (5) Di-pole or multi-pole antennas, (6) Stick antennas, (7) Beam-forming or phased-array element antennas, (8) Adaptive antennas, and (9) Omni directional antennas.

## 7. RFID Reader

RFID reader works as a central place for the RFID system. It reads tags data through the RFID antennas at a certain frequency [7, 9]. Basically, the reader is an electronic apparatus which produce and accept a radio signals [15]. The antennas contains an attached reader, the reader translates the tags radio signals through antenna, depending on the tags capacity [16]. The readers consist of a build-in anti-collision schemes and a single reader can operate on multiple frequencies. As a result, these readers are expected to collect or write data onto tag (in case) and pass to computer systems. For this purpose readers can be connected using RS-232, RS-485, USB cable as a wired options (called serial readers) and connect to the computer system. Also can use WiFi as wireless options which also known as network readers [8, 12]. Readers are electronic devices which can be used as standalone or be integrated with other devices and the following components/hardware into it [12].

Power for running reader, (2) Communication interface, (3) Microprocessor, (4) Channels, (5) Controller, (6) Receiver, (7) Transmitter, (8) Memory.

7.1 Tag Standards

Readers use near and far fields of methodology to communicate to the tag through its antennas [7]. If a tag wants to respond to the reader then the tag will need to receive energy and communicate with a reader. For example, passive tags use either one of the two following methods [7, 11].

**Near Fields:** Near field uses method similar to transformer, and employs inductive coupling of the tag to the magnetic field circulating around the reader antenna (see figure 8).

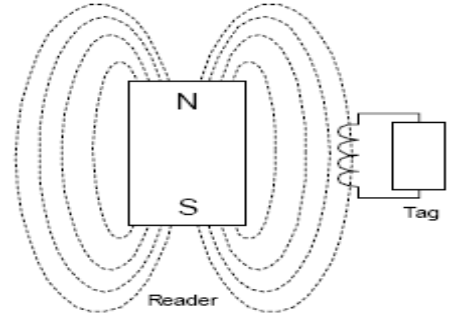

Fig. 8 RFID near field methodology [7]

**Far Field:** Far field uses method similar to radar, backscatter reflection by coupling with the electric field.

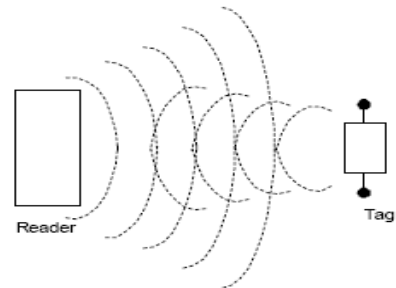

Fig. 9 RFID near field methodology [7]

The distinction between the RFID systems with far fields to the near fields is that the near fields use LF (lower frequency) and HF (higher frequency) bands [17, 18]. While RFID systems with far fields usually use longer read range UHF and microwave [17].

## 8. Advantages & Disadvantages of RFID System

Table 1: Comparison of RFID System

| Advantage | Disadvantage |
|---|---|
| High speed | Interference |
| Multipurpose and many format | High cost |
| Reduce man-power | Some materials may create signal problem |
| High accuracy | Overloaded reading (fail to read) |
| Complex duplication | |
| Multiple reading (tags) | |

## 9. Study Model

Our research aims to use "context based knowledge management" [4] to produce a model for mobile technology implementation within patients' movement processes. In order to do this we need to investigate RFID feasibility and integration with hospital information systems for improving healthcare [16]. This paper explores the fundamentals of RFID technology. That is the advantages, types, limitations and applications that will further help to develop a model for hospital application. The implementation of RFID in patient' flow is intended to improve healthcare especially in hospital settings. RFID technology can provide new capabilities as well as an efficient methods for several applications such as health care, access control, manage, store and analyze information inventory, business processes, and security controls through access to information [17]. The following section will describe the RFID applications [18] which will further help us to develop a model for mobile technology implementation within patients' movement processes.

## 10. Application Investigation

The hospital case study conducted by this research shows that there are objects which need to be considered when developing a model to represent patients' flow [4]. These objects are associated with location. The finite set of locations within the hospital will be captured through mobile technology in a live environment [16]. The following components have been observed in an overall picture of patients' movement [4]. However, the in depth investigation of each component is yet to be explored.

- The number of paramedical staff involved in patients' movement processes.
- The number of actions performed in patients' movement processes.
- The resources involved in an patients' movement processes.
- The finite number of locations used for patients' movement processes.
- The process of integrating patients' movement information with an existing IT infrastructure.

The system should enable the integration and optimization of resources while improving accuracy and minimizing patients' transition time leading to improvements in patients' services.

10.1 Type of RFID Applications

In previous section components are identified in hospital case. However, the detailed investigation is yet to be explored. In order to understand the benefits of RFID application in a hospital case, this paper explores general RFID applications shown in figure 10.

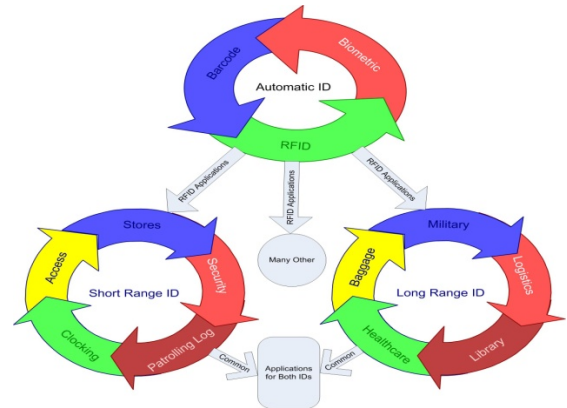

Figure 10. General RFID applications according to its capabilities

Object identification can be given through various ways such as barcode, biometric and RFID. RFID has two basic categories (short & long range). The **short range** applications need tags to be near reader, it is useful in various condition such as when a patient is required to come near the door/reader and only one person can get access (access control). The **long range** applications may not need tags that are closer to reader. Similar scenarios are successfully been used on various items in the warehouse (logistics) as shown in figure 9. The most common application is defined in figure 9 and is discussed in Section 10.2 together with other applications.

10.2 Healthcare Applications

RFID applications in healthcare could save important resources that can further contribute to better patient care. RFID applications could reduce the number of errors by tagging medical objects in the healthcare setting such as patients' files and medical equipment tracking in a timely manner. RFID further improves the situation for patients' care by integrating medical objects involved throughout the patients' care. RFID based timely information about the location of objects would increase the efficiency and effectiveness of paramedical staff leading to improved patients' experience [4, 16].

10.3 Security & Control Applications

RFID tags can be attached to the equipment/user personal/official belongings such as organization ID cards and vehicles. By applying RFID application in secure zones, not only permission can be granted to and revoke for the users/persons in particular zone but also record individual access and the length of their stay. It is also good for audit trial. These types of application consider time and flow carefully and as an aspect that is very important [19].

### 10.4 Patrolling Log Applications

RFID is also used for auditing and controlling security persons themselves. Application provides checkpoints for patrolling the security guards. Checkpoints are basically a RFID tag which security guard needs to scan during their sequential patrol through the reader. The reader maintains the record of the time and point at which the security guard swapped his card. This will not only help security firms administration to check the performance of its security guards but also used as a reference to track events. This application can also help to improve the patrolling process, e.g. through identifying the need to increase patrols or check points in a patrolled area.

### 10.5 Baggage Applications

Airline industries, package and delivery service lose a lot of money on lost or late delivery of baggage/packages. Handling large amount of packages from many places to various destinations on different routes can be very complex. In this scenario RFID application provide best resource management, effective operation and efficient transfer of packages. RFID helps to identify the packages, and provide records that can advice the industry on possible areas that may require some improvements. It also keeps customers informed about their packages.

### 10.6 Toll Road Applications

RFID applications make the toll collection/charging better with improved traffic flow, as cars/vehicles cannot pass through toll stations without stopping for payment. RFID is used to automatically identify the account holder and make faster transactions. This application helps to keep good traffic flow and to identify traffic patterns using data mining techniques that can inform the administration or decision support systems. For example, the information can be used to report the traffic conditions or to extend and develop future policies [19].

## 11. Conclusions

This study has identified and explained the nature of RFID technology evolution with respect to RFID applications. RFID technology will open new doors to make organisations, companies more secure, reliable, and accurate. The first part of this paper has explained and described the RFID technology and its components, and the second part has discussed the main considerations of RFID technology in terms of advantages and study model. The last part explores RFID technology applications. The paper considers RFID technology as a means to provide new capabilities and efficient methods for several applications. For example, in heathcare, access control, analyzing investory information, and business processes. RFID technology needs to develop its capability to be used with computing devices. This will allow businesses to get real potential benefits of RFID technology. This study facilitates adoption of location deduction technology (RFID) in a healthcare environment and shows the importance of the technology in a real scenario and application in connection with resource optimization and improving effectiveness. However, there is no doubt in the future that many companies and organisations will benefit from RFID technology.

**Future Work**

Our work continues to develop an enterprise architectural framework for managing contextual knowledge by exploiting object location deduction technologies such as RFID in healthcare processes that involve the movement of patients. Such a framework is intended to facilitate healthcare managers in adopting RFID for patient care resulting in improvements in clinical process management and healthcare services.

**Acknowledgments**

This research/project is funded by SaTH NHS Trust, UK and this RFID applications study is one of many objectives set in the larger project on "Context based knowledge management in healthcare".

**Kamran AHSAN** has an MSc in Mobile Computer Systems from Staffordshire University in Computer Science from University of Karachi. Kamran is a PhD researcher and lecturer in FCET (Faculty of Computing, Engineering and Technology) and, web researcher in Centre for Ageing and Mental Health, Staffordshire University, UK since 2005. He has published 10 research papers so far. He has involved in several research funding (UK) includes KTP, NHS Trust, Index Vouchers etc. He is Visiting Faculty at University of Karachi. He is a consultant to businesses in IT applications, software development and web tools. His research interests are mobile technology applications in healthcare including knowledge management.

**Hanifa Shah** is Professor of Information Systems and Director of the Centre for Information, Intelligence & Security Systems at Staffordshire University. She is also a Visiting Professor at Manchester University. Professor Shah has 30 years of experience as an IS/IT researcher and also practitioner. She is a Fellow of the British Computer Society and has led both UK Research Council (EPSRC) funded and industrially funded research projects. Her research has resulted in publications in international journals and conferences. She has research interests in a number of areas including the development of IS/IT systems, enterprise architecture, knowledge management, mobile technology based information systems, university-industry collaborations, learning through action research and the professional development of IS/IT practitioners in industry.

**Paul Kingston** is Professor of Health Sciences and Director of the Centre for Ageing and Mental Health at Staffordshire University. Professor Kingston has 34 years of experience as a Healthcare researcher and also practitioner. He has presented 101 conference papers in a number of different countries. He has co-authored or authored 7 books, and contributed 11 chapters. He has seen 33 journal articles published, produced 10 reports/ monographs, and 3 working papers. Additionally Paul has been involved in the development of training material in the area of adult protection and was a key founder of the charity Action on Elder Abuse. Furthermore he is a consulting editor for the Journal of Elder Abuse and Neglect. Over the last ten years Paul has developed a strong and respected track record for developing research activity. During ten years he was either in receipt of, co-managing, or leading awards totaling £1,785,633.